\documentclass[twocolumn,10pt]{revtex4}%
\usepackage[paperwidth=210mm,paperheight=297mm,centering,hmargin=2cm,vmargin=2.5cm]{geometry}
\usepackage{amsfonts}
\usepackage{amsmath}
\usepackage{amssymb}
\usepackage{bm}
\usepackage{graphicx}
\usepackage{color}
\usepackage{soul}
\usepackage{epsfig}%
\usepackage[dvipsnames]{xcolor} 
  \definecolor{bleu_cite}{RGB}{0,0,255}
\usepackage[colorlinks=true,allcolors = black,citecolor=bleu_cite,  ]{hyperref} 
\usepackage{natbib}

\begin{document}
\title{Defect-Driven Superfluid Crossover for Two-Dimensional Dipolar Excitons Trapped at Thermodynamic Equilibrium}

\author{Suzanne Dang$^1$, Romain Anankine$^{1}$, Carmen Gomez$^2$, Aristide Lema\^{i}tre$^2$,  Markus Holzmann$^3$ and Fran\c{c}ois Dubin$^{1}$} 
\affiliation{$^1$ Institut des Nanosciences de Paris, CNRS and Sorbonne University, 4 pl. Jussieu,
75005 Paris, France}
\affiliation{$^2$ Centre for Nanoscience
and Nanotechnology -- C2N, University Paris Saclay and CNRS, Route de
Nozay, 91460 Marcoussis, France}
\affiliation{$^3$ Univ. Grenoble Alpes, CNRS, LPMMC, 3800 Grenoble, France}

\begin{abstract}
We study ultra-cold dipolar excitons confined in a 10$\mu$m trap of a double GaAs quantum well. 
Based on the local density approximation, we unveil for the first time the equation of state
of excitons at pure thermodynamic equilibrium. In this regime we show that, below a critical temperature of about $1$ Kelvin, a superfluid forms in the inner region of the trap at a local exciton density $n \sim 2-3 \, 10^{10} \text{cm}^{-2}$, encircled by a more dilute and normal component in the outer rim of the trap. Remarkably, this spatial arrangement correlates directly with the concentration of defects in the exciton density which exhibits a sudden decrease at the onset of superfluidity, thus pointing towards an underlying  Berezinskii-Kosterlitz-Thouless mechanism.

\end{abstract}


\maketitle

In two dimensions, bosonic gases do not undergo a conventional Bose-Einstein condensation at finite temperatures \cite{Leggett_2006}. Instead, the Berezinskii-Kosterlitz-Thouless (BKT) theory \cite{Berenzinskii,KT} predicts that a topological phase transition may occur, from a normal to a superfluid phase where quasi-long range order is driven by the pairing between quantized vortices. The BKT crossover has originally been  observed with Helium films \cite{Reppy_1979} and more recently with ultra-cold atomic gases \cite{Dalibard_2006}, but its detection in the solid-state has remained more elusive. 

Superconducting films \cite{Halperin1979} constitute a natural candidate to explore the Berezinskii-Kosterlitz-Thouless crossover in the solid-state.  However, due to their electronic charge, Cooper pairs inevitably couple to  magnetic fluctuations, in- and out-of-plane, introducing further conceptual and practical challenges. Benefiting from most advanced epitaxial techniques, two-dimensional heterostructures based on GaAs quantum wells provide a concrete alternative to study neutral electron-hole pairs, strongly bound together by the attractive Coulomb potential and then forming a gas of bosonic quasiparticles interacting via their static dipole moment \cite{Combescot_book,Combescot_ROPP}. These  so called {\em excitons} actually offer a model system to probe exotic quantum phases of dipolar gases at thermal equilibrium \cite{Maciej_2011}, as shown by recent experiments reporting signatures of quantum coherence \cite{High_2012,Alloing_2014} and superfluidity \cite{Anankine_2017}. Characteristic features of a BKT transition in a driven-dissipative regime have also been observed for a polariton fluid \cite{Caputo_2018}, i.e. excitons coherently dressed by microcavity photons.

In this Letter, we show that dipolar excitons confined in coupled GaAs quantum wells exhibit the necessary degree of control to quantitatively study the phase diagram of neutral two-dimensional quasi-particles in a solid-state and at pure thermodynamic equilibrium. To this aim, we unveil for the first time the excitonic equation of state and then localise the crossover between normal and superfluid phases by the onset of spatial coherence which strongly correlates with the reduction of defects in the excitons photoluminescence. Our
quantitative study shows that exciton superfluidity in GaAs is characterised by an unusual four-component spin-structure and strong dipolar interactions. These ingredients open fascinating perspectives to explore exotic phases like supersolidity \cite{Li_2017,Maciej_2011}, or to approach the quantum phase transition at strong interactions where Bose condensation is suppressed even at zero temperature.

Originally introduced in the 1970's by Lozovik and Yudson \cite{Lozovik_76} dipolar excitons have since then received great attention in order to explore collective phenomena in semiconductors \cite{Butov_trap,Timofeev_trap,Holleitner_trap,Rapaport_trap,Beian_2017,Stern_2014}. Indeed, they offer a rather unique system to observe two-dimensional superfluidity in the limit of very strong inter-particle interactions \cite{Lozovik_QMC} and for intrinsically multi-component composite bosons, because they are made of $\pm1/2$ ($\pm$3/2) spin electron (hole) \cite{Combescot_2007,Combescot_2012,Combescot_2017}. The latter aspect is illustrated in Figure 1.a which shows that below a critical temperature of about 1K an exciton quasi-condensate is formed out of a dominant ($\sim$80$\%$) occupation of lowest energy and optically dark states, i.e. with a total "spin" equal to ($\pm$2), and a lower population of optically bright states with higher energy and a total "spin" equal to $(\pm$1). Such a condensate is not fragmented because fermion exchanges between excitons ensure a coherent coupling between the dark and bright parts \cite{Combescot_ROPP}, the latter one being crucial for our studies because it radiates the  photoluminescence unveiling quantum coherence of the many-body state \cite{Anankine_2017}. 


\begin{figure}\label{fig1}
\centerline{\includegraphics[width=.5\textwidth]{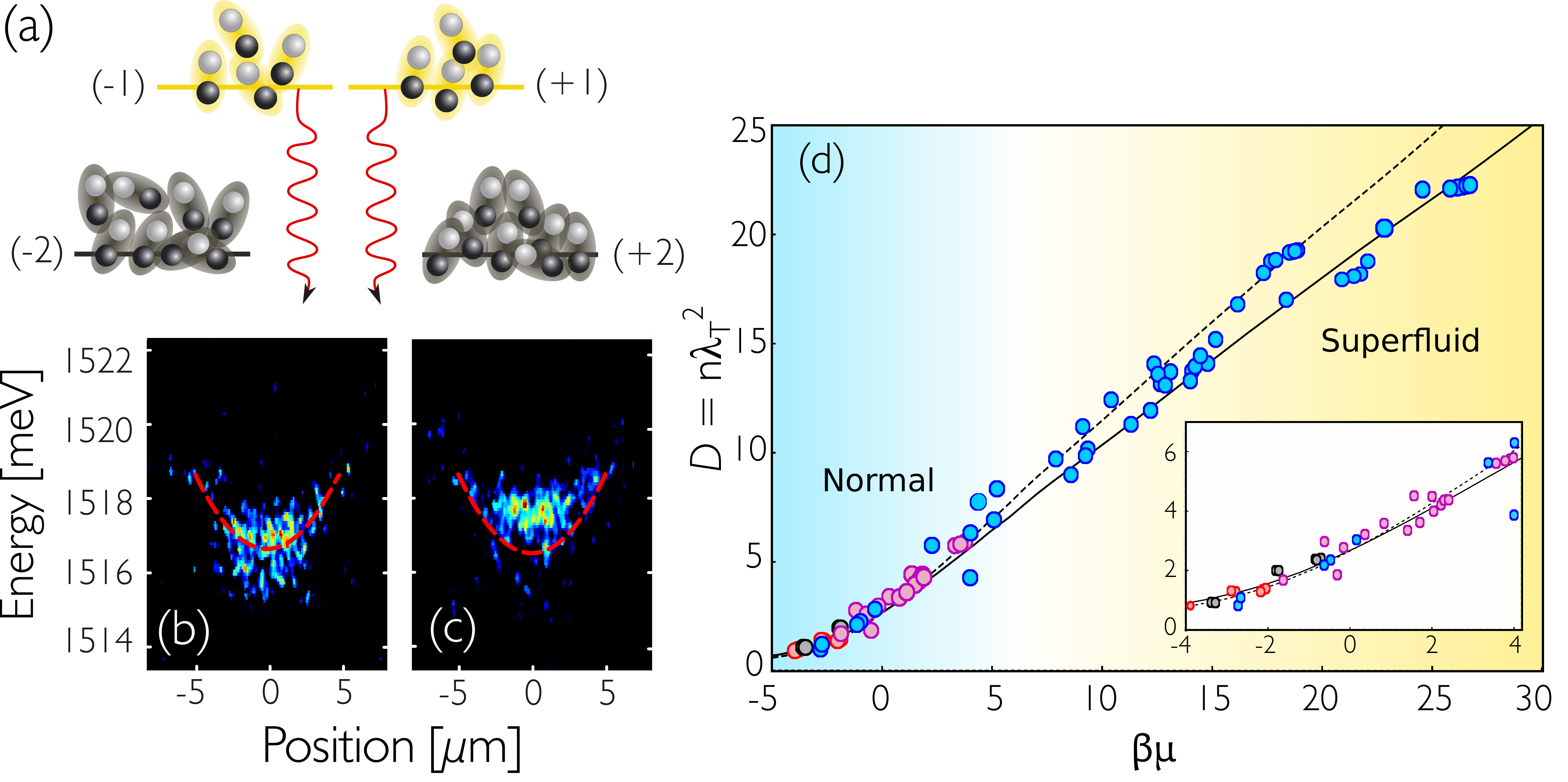}}
\caption{\textbf{Equation of state of the trapped gas} (a) Sketch of the four non-degenerate exciton spin states, bright ($\pm$1) and dark ($\pm$2). Below a critical temperature of about 1 K, excitonic condensation leads to a macroscopic occupation of dark states coherently coupled to a lower population of bright excitons radiating the analyzed photoluminescence (wavy red lines). (b-c) Photoluminescence resolved spectrally and spatially along the vertical axis of the electrostatic trap at T$_b$=340 mK. 350 ns after termination of the loading laser pulse we observe the energy profile of the trapping potential (b), well reproduced by a parabolic lineshape (dashed line). 150 ns after the loading laser pulse (c) the dispersion of the photoluminescence energy reflects the profile of the exciton density $n$ which is around 2.7$\cdot$10$^{10}$ cm$^{-2}$ at the trap center. (d) Phase space density $D$=$n$$\lambda_T^2$ as a function of the scaled chemical potential $\beta\mu$=$\mu/k_\mathrm{B}T_\mathrm{b}$. Experimental data are obtained by superposing density profiles measured at 3.5K (grey), 2.3K(red), 1.3K (pink) and 0.33 K (blue), up to a maximum density 2.7$\cdot$10$^{10}$ cm$^{-2}$ at the trap center. Variations expected from Monte-Carlo calculations are shown for $\tilde{g}$=6 and 7 (dotted and solid lines respectively). The inset presents a zoom of the dilute regime where we note a characteristic curvature of $D$($\beta\mu$).}
\end{figure}

In the following experiments we consider a GaAs bilayer heterostructure where electrons and holes are injected optically (Supplementary Informations). By imposing an electrical polarisation perpendicular to this device, we ensure that electrons and holes are each confined in a distinct layer. Due to Coulomb attraction, dipolar excitons are formed \cite{Lozovik_76}, and here confined in a 10 $\mu$m wide trap already studied in Ref. \cite{Anankine_2017,Beian_2017} (see also Supplementary Informations). The spatial separation enforced between electrons and holes constitutes a crucial ingredient. First, it provides
a long optical lifetime to dipolar excitons ($\gtrsim$100 ns see Ref. \cite{Beian_2017}) to reach thermodynamic equilibrium \cite{Ivanov_2004}. Further, it ensures that excitons experience repulsive dipolar interactions, their electric dipole d$\sim$12 C.nm being all aligned perpendicular to the bilayer. 

At a variable delay to a 100 ns long laser pulse loading a dense exciton cloud in the trap, we record the reemitted photoluminescence while the density is decreased due to radiative recombinations. For the 1.5 MHz repetition rate of our loading/detection sequence, note that a sufficient signal-to-noise ratio is typically reached for 5-10 seconds long acquisitions.  From the photoluminescence energy, E$_\mathrm{X}$, we extract the total exciton density, $n$(\textbf{r}), including all internal spin components, as well as the spatial profile of the confining 
potential, E$_\mathrm{t}$(\textbf{r}). Since E$_\mathrm{X}$(\textbf{r}) scales as (E$_\mathrm{t}$(\textbf{r})+$u_0$$n$(\textbf{r})) \cite{Ivanov_2010,Schindler_2008} where the latter term reflects the strength of repulsive dipolar interactions (Supplementary Informations), both, E$_\mathrm{t}$ and $n$, can be deduced directly in a single experiment by comparing  the spatial profile of E$_\mathrm{X}$ at different delays to the loading pulse. For a very long delay, the exciton density becomes sufficiently small so that its contribution in E$_\mathrm{X}$ is neglegible. Figure 1.b presents the profile of the trapping potential thus measured at a bath temperature T$_\mathrm{b}$= 340 mK, whereas Fig.1.c shows the profile of E$_\mathrm{X}$ when about 2.75 10$^{10}$ cm$^{-2}$ excitons are confined at the trapping center.

In our studies the photoluminescence is spectrally narrowband (around 500 $\mu$eV width \cite{Beian_2017,Anankine_coherence}) which allows us to accurately extract  density profiles across the trap, the blueshift of the photoluminescence being of at most 1 meV.
Within the local density approximation, i.e. associating the local density, $n$(\textbf{r}), to the
local chemical potential, $\mu$(\textbf{r})=$\mu_0$-E$_\mathrm{t}$(\textbf{r}), 
we can extract the excitons equation of state, $n$($\mu$, $T_\mathrm{b}$), exploring various delays to the loading pulse and bath temperatures.
While these density profiles are all different, Figure 1.d  shows that the phase space density, $D$=$n$$\lambda^2_\mathrm{T}$, collapses remarkably well to a single curve in scaled units $\beta\mu$=$\mu/k_\mathrm{B}T_\mathrm{b}$,  $\lambda_\mathrm{T}$=h/$\sqrt{2\pi mk_\mathrm{B}T_\mathrm{b}}$ being the de Broglie thermal wavelength with $m$ the excitons effective mass. This scale invariance of the excitons phase space density, $D$, is obtained for a set of 30 experiments realised at temperatures, $T_\mathrm{b}$, ranging from 0.34 to 3.5 K, and for a maximum density about 2.75 10$^{10}$ cm$^{-2}$ at the trap center. Such scale invariance of the equation of state has been observed in quasi-two-dimensional cold atomic gases \cite{Hung_2011,Yefsah_2011}, our measurements providing the counterpart in a quite
distinguished solid state system thus revealing for the first time the thermodynamic equilibrium for excitons.

The equation of state is extremely valuable to microscopically describe the excitons many-body state, a formidable theoretical challenge \cite{Combescot_book}.
Considering excitons as point-like bosons,
the scattering between them is dominated by their dipolar interaction, which at two dimensions is sufficiently short-ranged to be modelled by a structureless contact interaction characterized by
a dimensionless coupling constant $\tilde{g}$.
Based on the effective Hamiltonian described in Ref. \cite{Combescot_2017} and considering all four accessible excitonic spin states, we have performed classical field Monte Carlo calculations to quantify the universal behavior in the quantum degenerate regime \cite{Prokofev2002}, adding
non-universal quantum corrections within mean-field to obtain the full equation of state \cite{HCK}.
For $\tilde{g}$$\sim$6, this effective low energy description reproduces quantitatively our 
observations, regardless the precise values set for the few $\mu$eV \cite{Lavallard}
energy splitting between the different spin states.
Remarkably, this amplitude for $\tilde{g}$ agrees closely with first principle estimations (Supplementary Informations) and independent theoretical treatments \cite{Lozovik_QMC}. 
This shows that dipolar excitons provide a new testbed for the strong interaction regime of a gaseous bosonic phase, i.e. with $\tilde{g}$ of order one; so far this regime was only reached using Feshbach resonances in atomic systems \cite{Zoran,Boettcher_2016}, or with liquid Helium films \cite{Reppy_1979}.
We note that in both of these systems, our effective description accurately predicts the Kosterlitz-Thouless transition within a few percent compared to full quantum Monte Carlo calculations \cite{HCK,Pilati2008}. 

Our theoretical model predicts the occurence of a BKT transition above $D_c \approx 8$.
For the experiments displayed in Fig.1.d, this critical density is only reached for T$_\mathrm{b}\leq$1.3 K, even at the center of the trap where the density is the largest. Importantly, this threshold temperature matches the one measured for the emergence of quasi long-range order under identical experimental conditions \cite{Anankine_2017}. Moreover, Fig.2.a shows that $D\geq$$D_c$ is only 
reached when the distance to the center of the trap $||\textbf{r}||$ is less than about 3$\mu$m at T$_\mathrm{b}$=340 mK. We thus expect that the superfluid state formed in the central region is encircled by a ring-shaped lower density and normal component.

\begin{figure}\label{fig2}
\centerline{\includegraphics[width=.5\textwidth]{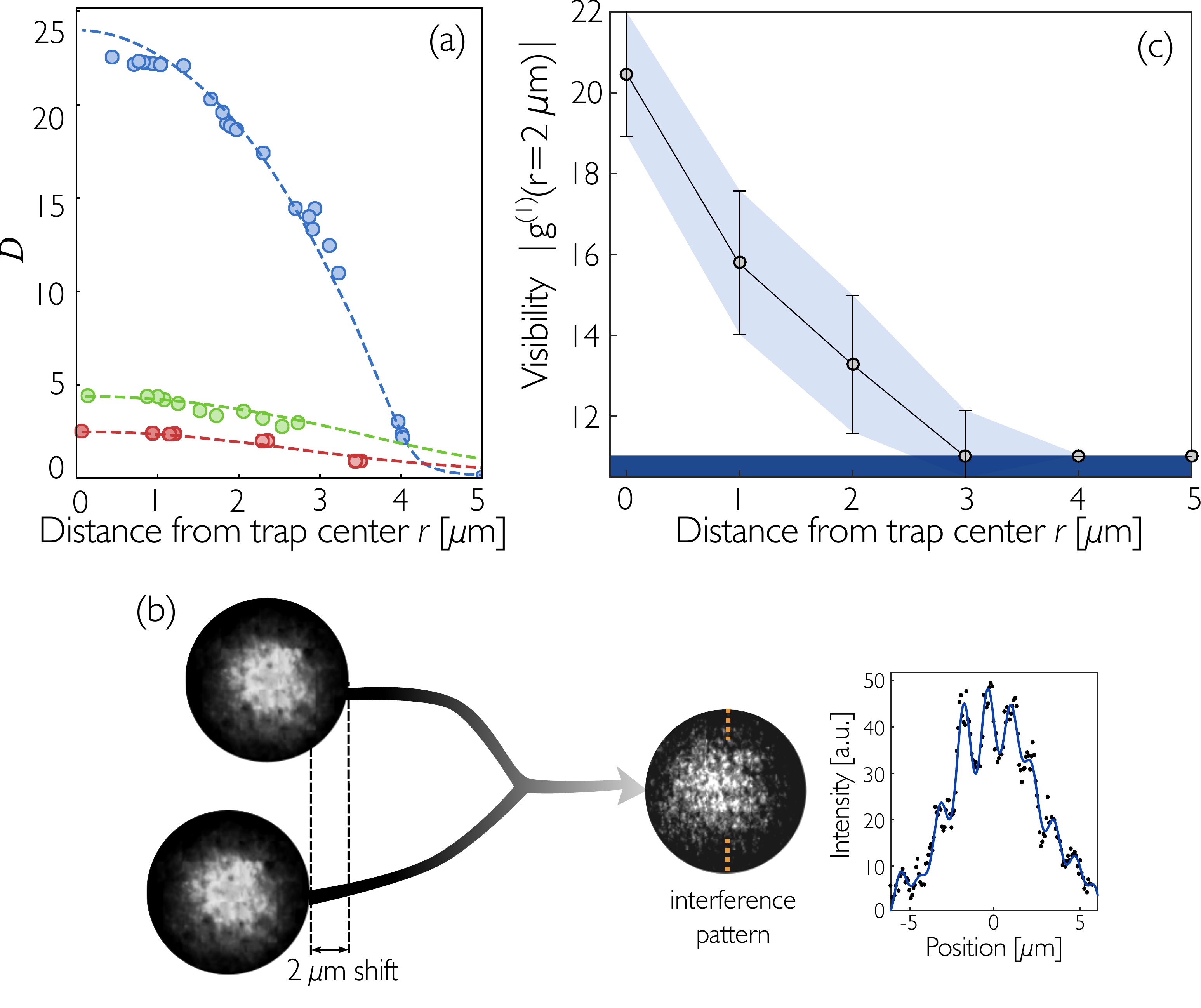}}
\caption{\textbf{Spatial coherence at T$_\mathrm{b}$=340 mK} (a) Exciton phase space density $D$ measured across the trap at 340 mK (blue), 1.3K (green) and 2.5K (red) and for a central density of about 2.7$\cdot$10$^{10}$ cm$^{-2}$. Dashed lines represent the profiles expected by Monte-Carlo calculations at each bath temperature. (b) The spatial coherence of the quasi-condensate is assessed by splitting the photoluminescence in two equal parts, recombined after a lateral shift of 2 $\mu$m has been introduced. The interference visibility $V$ is measured by the amplitude modulation along the vertical direction, as shown by the right panel for a profile taken in the inner part of the trap. (c) Variation of $V$ as a function of the distance to the trap center, for a statistical average of 90 experiments where $n$($||\textbf{r}||$=0)$\sim$2.7$\cdot$10$^{10}$ cm$^{-2}$ at T$_\mathrm{b}$=340 mK, i.e. for the same experimental conditions as in Fig.1.c. The minimum contrast of 11$\%$ is given by the signal to noise ratio of our measurements.}
\end{figure}

To experimentally confirm the quasi-long range order of excitons in the inner of the trap, we performed interferometric measurements where the photoluminescence is divided in two parts which are recombined after introducing a 2 $\mu$m lateral shift, i.e. about 10 times the classical limit set by $\lambda_\mathrm{T}$ (Fig.2.b).  Thus, we quantify the excitonic first-order correlation function $|g^{(1)}|$ that is given by the interference visibility $V$. Indeed, for our studies the exciton-photon coupling is linear  \cite{Combescot_book} so that the coherence of the quasi-condensate is imprinted in the photoluminescence radiated by its bright part (Fig.1.a).  

Fig.2.c quantifies the interference contrast across the trap at T$_\mathrm{b}$=340 mK and for the same experimental conditions as in Fig.1.c, i.e. for a density of about 2.75 10$^{10}$ cm$^{-2}$ at the center. Our experimental conditions varying slightly from one measurement to the following one (Supplementary Informations), we statistically averaged the results of an ensemble of 90 realisations to reach relevant conclusions. Then, we note that at the trap center $V$  is about half its value for zero lateral shift, manifesting directly the non-classical nature of the emission. The contrast then decreases with increasing distance to the center up to $||\textbf{r}||\simeq$3 $\mu$m where we reach the magnitude set by the signal-to-noise S/N ratio of our experiments.
Further, we verified that $V$ does not exceed the value set by the S/N ratio throughout the all 
trap for T$_\mathrm{b}\gtrsim$1.3 K, so that coherence neither develops in this higher temperature range, in agreement with the predicted critical phase space density $D_c$.

We now show that the sharp cross-over in the excitons coherence can be independently
characterized by mapping out the spatial profile of local defects in the
photoluminescence pattern. The remarkable correlation between both features reenforces the picture of a defect-driven phase transition in our system. 

\begin{figure}\label{fig3}
\centerline{\includegraphics[width=.5\textwidth]{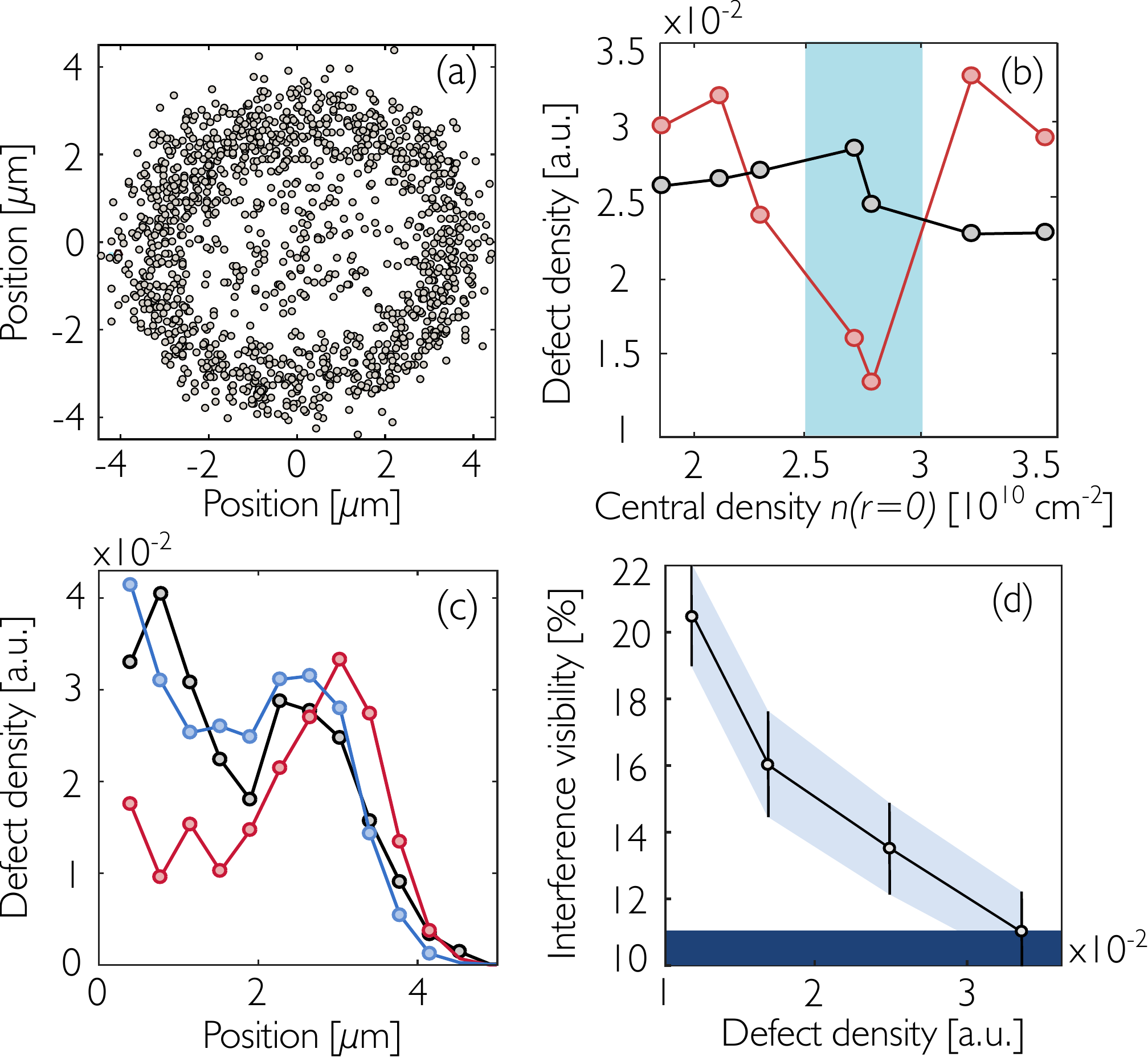}}
\caption{\textbf{Photoluminescence defects at T$_\mathrm{b}$=340 mK} (a) Cartography of photoluminescence defects measured for 60 experiments where $n$($||\textbf{r}||$=0)$\sim$ 2.7$\cdot$10$^{10}$ cm$^{-2}$, i.e. for the same experimental conditions as for Fig.1.c and Fig.2.c. (b) Defects density $\mathcal{P}$ as a function of $n$($||\textbf{r}||$=0), in the inner region of the trap ($||\textbf{r}||\leq$1.5$\mu$m) in red and in the outer region of the trap (1.5$\leq||\textbf{r}||\leq$ 3$\mu$m) in black. The blue area underlines the density range where quantum spatial coherence is resolved. (c) Spatial profile of the defects density for $n$($||\textbf{r}||$=0)$\sim$ 2.7$\cdot$10$^{10}$ cm$^{-2}$ (red), 1.8$\cdot$10$^{10}$ cm$^{-2}$ (black) and 3.5$\cdot$10$^{10}$ cm$^{-2}$ (blue). (d) Variation of the spatial interference contrast as a function of the defects density for $n$($||\textbf{r}||$=0)$\sim$ 2.7$\cdot$10$^{10}$ cm$^{-2}$.} 
\end{figure}

In Fig.2.b we recognize strong local fluctuations in the photoluminescence pattern radiated from the trap. Such defects are due to local electrostatic fluctuations of the trapping potential and let us then stress that their positions vary within few seconds during measurements performed under fixed experimental conditions, defects being then randomly distributed over space \cite{Anankine_2017}. Since the bare electrostatic noise of the order of a few hundreds of $\mu$eV is expected to be effectively screened \cite{Ivanov_2004} by dipolar exciton-exciton repulsions ($\sim$1 meV), the occurence of such strong density fluctuations  is rather surprising. In Figure 3.a we display a cartography of  the defects detected in the photoluminescence at  T$_\mathrm{b}$=340 mK and $n$($||\textbf{r}||$=0)$\sim$ 2.75 10$^{10}$ cm$^{-2}$, i.e. for the regime discussed in Fig.2.c (see Supplementary Informations). The positions of defects varying between successive measurements, Fig.3.a is obtained for a statistical ensemble of 60 realisations and shows that  a small concentration of defects is present in the central region of the trap ($||\textbf{r}||$$\lesssim$2$\mu$m) compared to the outer rim where most of the defects are located. Importantly, this inhomogeneous distribution only occurs inside a narrow range, i.e. for n($||\textbf{r}||$=0)$\sim$ 2.5-3 10$^{10}$ cm$^{-2}$. Indeed, the defect density $\mathcal{P}$ increases sharply in the inner of the trap at lower and higher densities (Fig.3.b-c). In the latter case the invariance of the equation of state is actually violated such that this regime is not considered in Fig.1d. The correlation between the defect concentration and the degree of quantum coherence is directly evidenced in Fig.3.d. There, we show that the interference visibility $V$ decreases monotonously with increasing $\mathcal{P}$ inside the superfluid region. Quantum coherence emerges then only below a threshold of density fluctuations, similarly to observations with ultracold atomic gases \cite{Dalibard_2006}.

\begin{figure}\label{fig4}
\centerline{\includegraphics[width=.5\textwidth]{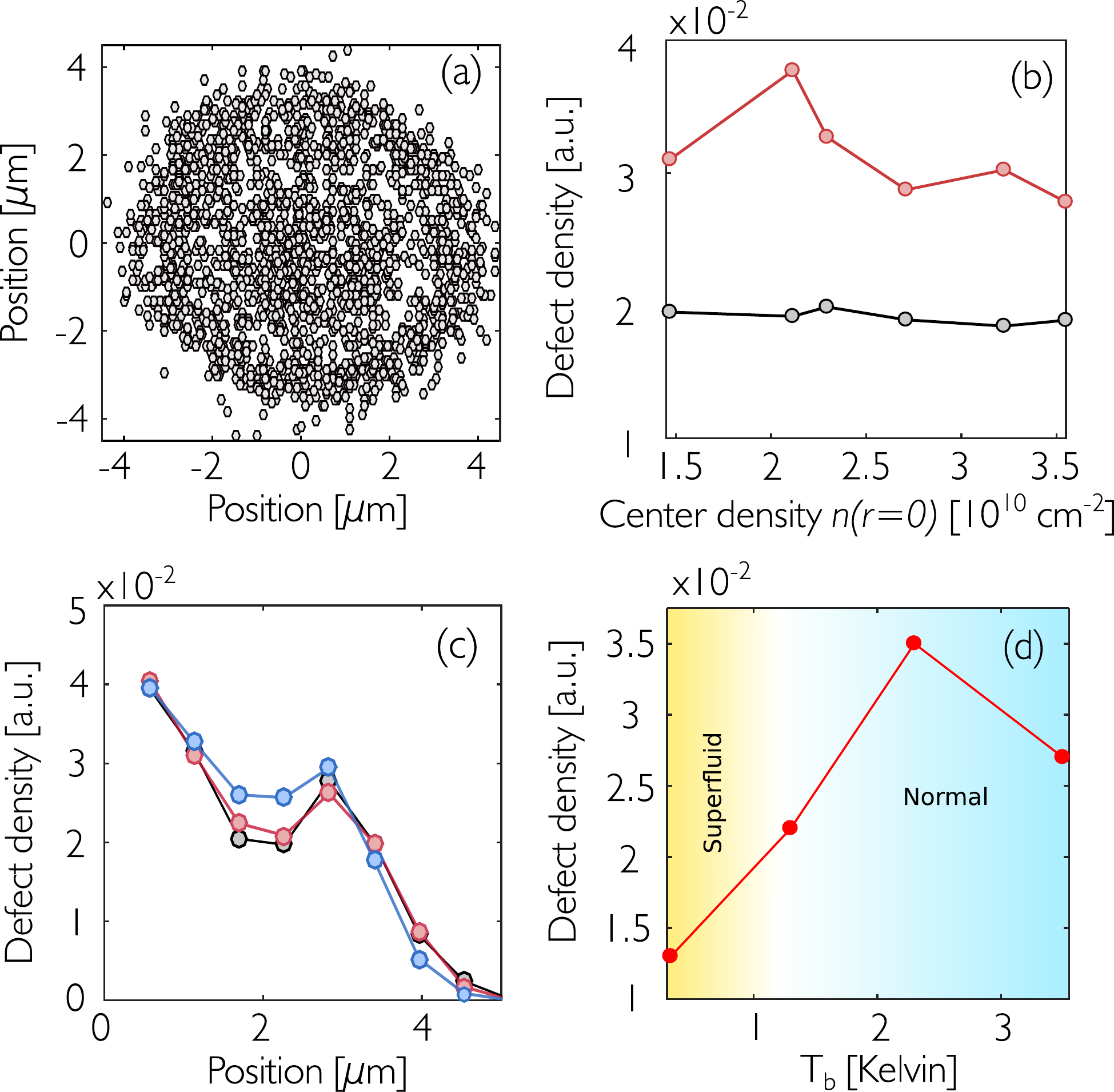}}
\caption{\textbf{Density fluctuations vs. bath temperature} (a) Cartography of photoluminescence defects measured for 60 experiments where $n$($||\textbf{r}||$=0)$\sim$ 2.7$\cdot$10$^{10}$ cm$^{-2}$ at 2.3K. (b) Defects density at T$_\mathrm{b}$=2.3K averaged in the inner region of the trap ($||\textbf{r}||\leq$1.5$\mu$m) in red and in the outer region of the trap (1.5$\leq||\textbf{r}||\leq$ 3$\mu$m)  in black, as a function of the central density $n$($||\textbf{r}||$=0). (c) Defects density $\mathcal{P}$ resolved across the trap at T$_\mathrm{b}$=2.3K, for $n$($||\textbf{r}||$=0)$\sim$ 2.7$\cdot$10$^{10}$ cm$^{-2}$ (red), 1.5$\cdot$10$^{10}$ cm$^{-2}$ (black) and 3.5$\cdot$10$^{10}$ cm$^{-2}$ (blue). (c) . Measurements correspond to an average of 60 experiments at every density. (d) Variation of the defects density  averaged in the inner region of the trap ($||\textbf{r}||\leq$1.5$\mu$m) as a function of the bath temperature. For every measurement the central exciton density is kept at 2.7$\cdot$10$^{10}$ cm$^{-2}$.} 
\end{figure}

To further quantify the intimate relation between the concentration of photoluminescence defects and the superfluid crossover, we study in Figure 4 the distribution of defects at higher bath temperatures. First, we show in Fig.4.a-c that $\mathcal{P}$ varies monotonously at T$_\mathrm{b}$=2.3 K from the inner to the outer rim of the trap. In contrast to the sub-Kelvin regime,  $\mathcal{P}$ does not depend on the exciton density, neither in the center nor in the outer rim of the trap (Fig.4.b-c). In the latter region, $\mathcal{P}$ actually does not significantly vary with the bath temperature. However, for densities around n($||\textbf{r}||$=0)$\sim$2.75 10$^{10}$ cm$^{-2}$ where  
quantum coherence emerges below about 1.3 K \cite{Anankine_2017}, Fig.4.d highlights
 that the defect concentration in the inner region is decreased by over two-fold
between 2 and 0.34 K. This reveals that upon a temperature decrease, quasi long-range order is established together with a strong reduction of photoluminescence fluctuations. 

It seems natural to attribute photoluminescence defects to quantized vortices whose proliferation drives the
transition from the superfluid to normal phase according to the BKT scenario.
This interpretation is actually well supported by previous experiments \cite{Anankine_2017} which could associate photoluminescence defects to phase singularities in the radiation of the trap. However, we want to point out that the spatial extension of quantized vortices is expected to be around 25 nm, e.g. of order of the mean-field coherence length $\xi$=1/$\sqrt{2\tilde{g}n}$, so that the direct observation of free vortices is well below our optical resolution ($\sim$1$\mu$m). The detection of phase singularities is therefore only possible when quantized vortices
are pinned by a weak potential fluctuations, such that a 2$\pi$ phase shift is revealed around a considered defect. Let us then note that quantum Monte Carlo calculations \cite{Carleo2013} indicate that the BKT transition is robust against local disorder potentials, up to amplitudes of the order of the chemical potential which is consistent with our experimental observations.

To conclude, let us note that even deep in the superfluid phase
at T$_\mathrm{b}$=340 mK and n($||\textbf{r}||$=0)$\sim$2.75 10$^{10}$ cm$^{-2}$, where
the phase space density  $D \sim 20$ exceeds more than twice the critical value $D_c$,
a noticeable concentration of defects remains visible. 
As vortex pairs are thermally activated, their concentration rapidly decreases as the phase
space density exceeds $D_c$ in conventional single component superfluids.
Although we cannot exclude electromagnetic fluctuations as a source for this atypical behaviour,
 it is likely that in the superfluid defects result from the almost degenerate four exciton spin components, since bright states are still occupied by un-condensed excitons corresponding to aound 20$\%$ of the total population at T$_\mathrm{b}$=340 mK \cite{Anankine_2017}. This conclusion is directly supported by our Monte Carlo calculations showing that quasi-condensation occurs only in the lowest energy four-component spin state, whereas the populations of the incoherent higher energy spin components saturate at a value sufficiently high to support and activate vortices in the system.
 
 \textbf{Acknowledgements:} The authors are grateful to  M. Combescot, X. Xu, J. Dalibard, M. Lewenstein, R. Combescot and A. Leggett for a critical reading of the manuscript. We would also like to thank K. Merghem and E. Cambril for their contribution to the sample processing. Our work has been financially supported by the projects XBEC (EU-FP7-CIG) and by OBELIX from the french Agency for Research (ANR-15-CE30-0020). 
 
 \vspace{3cm}
 
 
 \Large{\textbf{Supplementary Informations}
 
 \normalsize
 
 \section{Sample structure and experimental procedures}

We study the same electrostatic trap as in Refs. \cite{Beian_2017,Anankine_2017}. Let us then briefly recall the main properties of this device.

The electrostatic trap is made by a field-effect device of about 1 $\mu$m thickness and embedding two 8 nm wide GaAs quantum wells separated by a 4 nm Al$_{.3}$Ga$_{.7}$As barrier. The two quantum wells are positioned 900 nm below the surface of the heterostructure where 2 semi-transparent gate electrodes are deposited after electronic lithography. On the other hand, 150 nm below the quantum wells, a Si-doped GaAs layer serves as back contact for the field-effect heterostructure. 
\vspace{.5cm}

\centerline{\includegraphics[width=.5\textwidth]{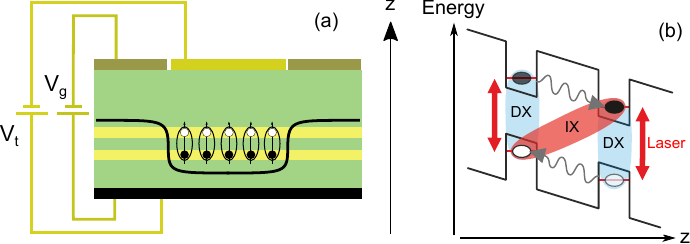}}
\textit{\textbf{Fig. S1}: (a) Our 10$\mu$m trap is engineered by a disk electrode surrounded by a larger guard electrode. Applying a stronger bias onto the central electrode yields a trapping potential for dipolar excitons that behave as high-field seekers. (b) Electron-hole pairs are injected directly in the electrostatic trap by means of pulsed laser excitation at resonance with the absorption of the direct excitons absorption (DX). Dipolar excitons (IX) are formed once electronic carriers have tunnelled towards their minimum energy states (wavy gray lines).}
\vspace{.5cm}

As illustrated in Fig.S1, the 10$\mu$m wide trap is implemented using a central disk-shape electrode separated by 200 nm from an outer guard gate. Spatially indirect (or dipolar) excitons, e.g. made of an electron confined in the bottom quantum well bound to a hole confined in the top quantum well, are attracted towards the regions where the electric field applied perpendicular to the quantum wells is the largest. Thus, by applying onto the disk electrode  a stronger potential  than onto the guard we directly realize a two-dimensional trap \cite{Beian_2017}. In our experiments, we actually make use of the potential rectification between the two gates in order to define the trapping area \cite{Anankine_2017}. We follow this approach to minimize the amplitude of the electric-field in the plane of the quantum wells.

As in previous studies \cite{Beian_2017,Anankine_2017} we used a 100 ns long laser excitation in order to inject optically electrons and holes in the two quantum wells, dipolar excitons being created once carriers have tunnelled towards their minimum energy states (Fig.S1.b). The laser is set resonant with the direct excitonic absorption of the quantum wells, such that the density of optically injected excess free carriers is minimized, these being directly linked to the amplitude of a transient photo-current  ($\sim$50 pA DC). Let us note that the laser beam covers most of the trap area and that we typically impose a dead time of around 100 ns before analyzing the photoluminesence radiated from the trap. Thus, we ensure that our measurements are restricted to the regime where the photo-current is completely evacuated. 

\centerline{\includegraphics[width=.35\textwidth,angle=90]{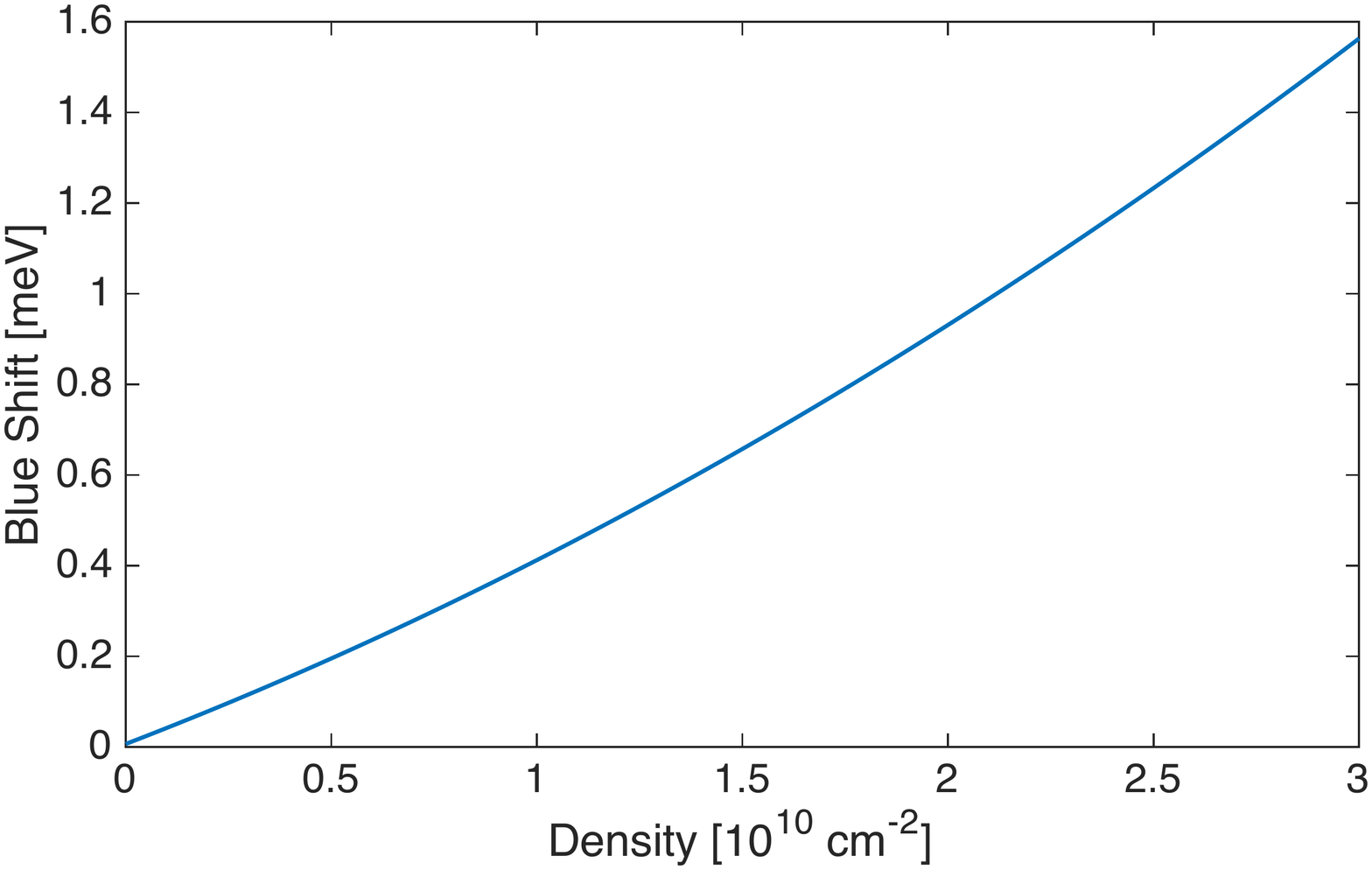}}
\textit{\textbf{Fig. S2}: Blue shift energy of the photoluminescence as a function of the density of dipolar excitons.}
\vspace{.5cm}

In the main manuscript, we report experiments where the photoluminescence is detected in a 10 ns long interval, separated by a delay $\tau$ set between 100 to 350 ns after extinction of the loading laser pulse, our experimental sequence being repeated at a frequency equal to 1.5 MHz during typically 5 to 10 seconds. Due to radiative recombinations, while $\tau$ increases the exciton density is decreased in the trap. In the main text, we underline that the photoluminescence energy $E_X$ is used to infer the density  $n_X$. For that purpose, we precisely monitor the blue-shift energy, that is the difference between $E_X$ at a given delay to its value for $\tau$=350 ns, i.e. when the trapped gas is sufficiently dilute so that the density correction to the photoluminescence energy is negligible. 

To calibrate the exciton density we follow the theoretical approach introduced by Ivanov, Zimmerman and co-workers \cite{Ivanov_2010} which explicitly takes into account the impact of dipolar repulsions between excitons. Moreover, note that we neglect quantum corrections to the photoluminescence energy.  In Fig. S2 we show the resulting variation of the blue shift energy as a function of the exciton density for our physical parameters, e.g. for a 12 nm spatial separation between electrons and holes. Using this calibration we deduce that $n_X\sim$ 2.7 10$^{10}$ cm$^{-2}$ for $\tau$=150 ns which corresponds to the regime where we find superfluid signatures in the sub-Kelvin temperature range.

\section{Measurements of the excitons degree of spatial coherence}

\centerline{\includegraphics[width=.5\textwidth]{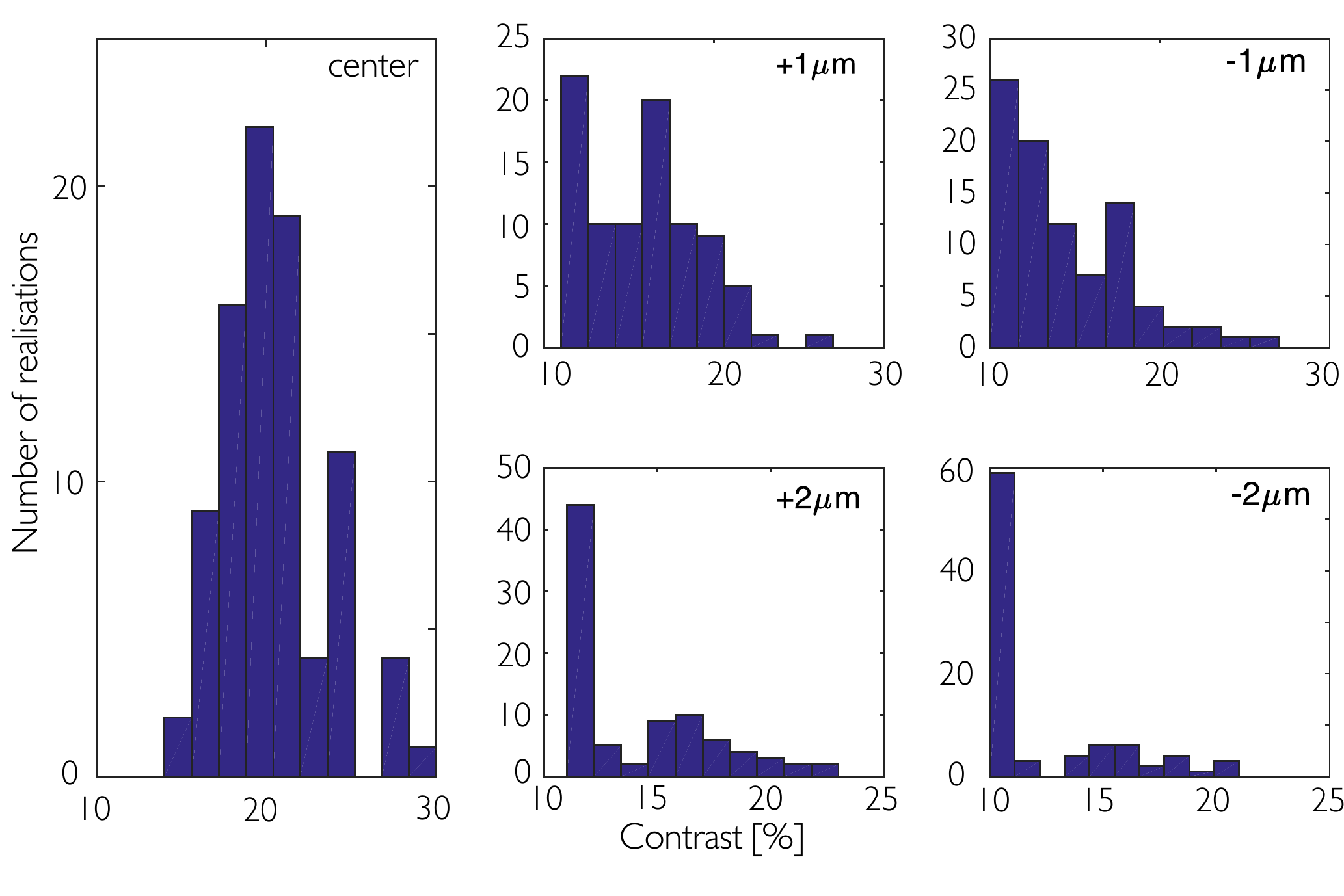}}
\textit{\textbf{Fig. S3}: Interference contrasts measured at the center of the trap and, and at a distance of $\pm$1 $\mu$m and $\pm$2 $\mu$m to the center. Experimental data are obtained by considering 90 experiments, all performed for an exciton density at the center of the trap of about 2.7 10$^{10}$ cm$^{-2}$ at T$_\mathrm{b}$=340 mK.}\newline

The spatial coherence of the photoluminescence is measured using a Mach-Zehnder interferometer where the output of one arm is displaced horizontally by 2 $\mu$m compared to the other arm. We also introduce a vertical tilt angle between the output of the two arms in order to set horizontal fringes in the resulting interference pattern. The same approach was already used in the experiments reported in Refs. \cite{Anankine_2017, Alloing_2014}.

As underlined in the main text, the data displayed in Fig. 2.b is obtained by statistically averaging the contrasts measured for a set of interference patterns, all acquired for $\tau$=150 ns at T$_\mathrm{b}$=340 mK i.e. for the experimental conditions where we observe quantum coherence. This statistical averaging was motivated by the fluctuations of the excitons electrostatic confinement, such that we do not measure the exact same interference pattern between successive experiments performed under fixed conditions \cite{Anankine_2017}. Fig.S3 shows the resulting distribution of interference contrasts as a function of the distance to the trapping center. Note that a minimum value equal to 11$\%$ is assigned to the profiles where an interference signal is not identified. This value is given by the signal to noise ratio in our experiments and we obtained the results shown in Fig.2.c of the main text after averaging the positions symmetric to the center of the trap. Moreover, let us underline that the interference contrasts are always evaluated without subtracting any background.

\section{Defects in the photoluminescence radiated from the trap}

The density fluctuations discussed in Fig.3-4 of the main text correspond to the positions where the intensity in the real image of the photoluminescence is decreased locally by at least 20$\%$. For the signal to noise ratio of our experiments, such intensity loss corresponds to a statistical deviation greater than 2$\sigma$. To obtain the results discussed in the main text, we systematically consider an ensemble of 60 realisations from which we extract the coordinates of the photoluminescence defects, for each delay to the termination of the loading laser pulse and each bath temperature. Thus,  we statistically compute the density of probability to find photoluminescence defects as a function of the distance to the trapping center.

To illustrate directly how the density of photoluminescence defects varies in the inner region of the trap, we present in Fig.S4 characteristic maps of the defects positions measured while the density in the trap is varied. At T$_\mathrm{b}$=340 mK, we note directly in the left panels of Fig.S4 that the concentration of defects is minimized around the center of the trap only for a narrow range of densities (middle panel). On the other hand, at T$_\mathrm{b}$=2.3K (right panels of Fig.S4) this behaviour is no longer observed and the density of defects does not vary significantly with the exciton density in the inner of the trap.

\vspace{.5cm}

\centerline{\includegraphics[width=.4\textwidth]{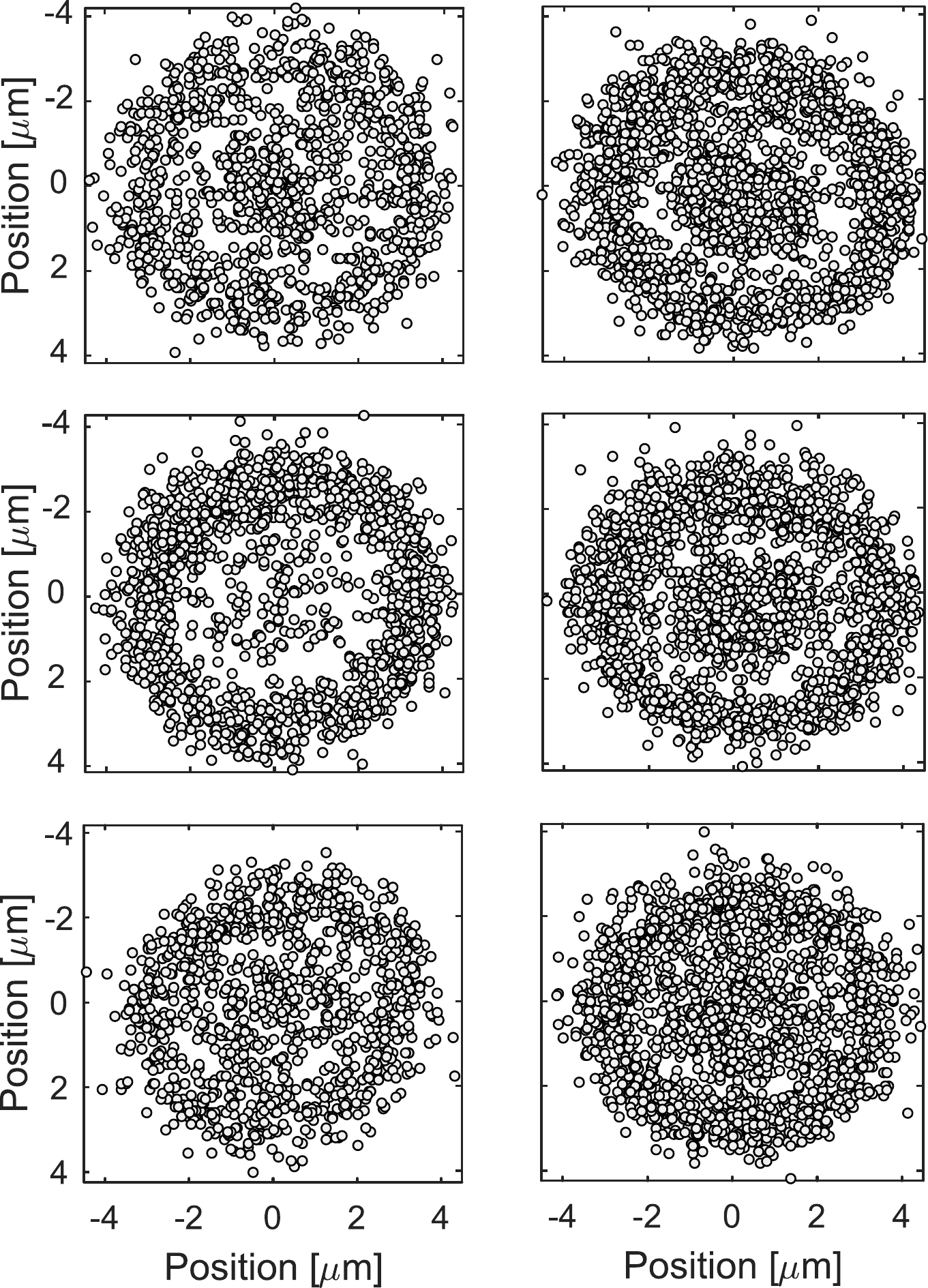}}
\textit{\textbf{Fig. S4}: (Left panels) Positions of the photoluminescence defects detected for 60 experiments performed T$_\mathrm{b}$=340 mK, with a central density $n$($||\textbf{r}||$=0) $\sim$ 1.5 (top), 2.7 (middle) and 3.5 10$^{10}$ cm$^{-2}$ (bottom). (Right panels) Cartography of the photoluminescence defects detected for the same experimental conditions but for T$_\mathrm{b}$=2.3 K. }

\section{Modelling dipolar interactions by a contact potential}

The Schr{\"o}dinger equation for the relative motion of two excitons interacting via a dipolar potential 
reads
\begin{equation}
\left[ -\frac{\hbar^2 \nabla^2 }{m_X} + \frac{d^2 e^2}{r^3} - \varepsilon \right] \psi_\varepsilon=0
\end{equation}
where $m_X$ denotes the exciton effective mass, $d$ the length of their dipole moment and  $e$ the electron charge. At low energies,
the short-distance behavior is dominated by the zero energy solution, $\psi_{\varepsilon=0}$, 
\begin{equation}
\Psi_0(r) \sim K_0(2 \sqrt{r_0/r})
\end{equation}
where $K_0(z)$ is the modified Bessel function, 
and $r_0=d^2/a_X$ introduces a characteristic length scale with $a_X=\hbar^2/m_Xe^2$ the exciton Bohr radius.

At distances $r$ large compared to $r_0$, we then have
\begin{equation}
\psi_0(r) \sim - \log \frac{e^{2 \gamma}}{r/r_0}
\end{equation}
where $\gamma\sim0.5772$ is the Euler-Mascheroni constant. We can then match this large distance asymptotic behaviour with the s-wave radial part of the
general solution of the free (non-interacting) Schr{\"o}dinger equation, of
energy $\varepsilon=\hbar^2 k^2/m_X$, which reads
\begin{equation}
R_0(kr) \sim \cos \delta_0(kr) J_0(kr)-\sin \delta_0(k) Y_0(kr)
\end{equation}
the phase-shifts $\delta_0(k)$ being determined in the region $r_0 \ll r \ll 1/k$.
By noting that the same asymptotic behaviour can be obtained from a 
contact pseudo-potential of dimensionless, energy-dependent, coupling strength 
$\tilde{g}=-4 \tan \delta_0(k)$, we obtain for the dipolar interaction
\begin{equation}
\tilde{g}=\frac{2 \pi}{\log \frac{2 e^{-3 \gamma}}{kr_0}}
\end{equation}
which exhibits a logarithmic energy dependence, expected for two-dimensional scattering.

For dipolar excitons, we have $a_X\simeq 20$nm and $d \simeq 12$nm which leads to $r_0 \simeq 7$nm. This value is considerably smaller than both the thermal wave length at the considered temperatures and the typical spatial separation between excitons. Therefore, it is reasonable that the dipolar interaction between excitons is fairly well described by an effective contact potential.

Using $k\sim (m_X k_B T)^{1/2}$ in the thermal region, from Eq.(5) we deduce that $\tilde{g} \sim 4$. For such a
large coupling constant, the energy dependence, although logarithmic, is no longer negligible.
In contrast to the weakly interacting situation, $\tilde{g} \ll 1$, we therefore expect that the
scale invariance is ultimately broken, though it may hold approximately within the
experimental uncertainty in a fairly large parameter regime.

Let us moreover note that although modelling the many-body interaction between excitons via a contact potential is  reasonable, the value obtained from the experimental equation of state, $\tilde{g} \sim 6$, should be regarded as an effective coupling constant. Indeed, since we neglected intrinsic quantum fluctuations in our model calculation of the equation of state, $\tilde{g} \sim 6$ probably overestimates the bare two-body interaction scattering  at this energy scale. 

Finally, we would like to mention that weak fluctuations of the trapping potential mainly introduce small shifts of the chemical potential, while the universal behavior of Kosterlitz-Thouless transition is expected to be
robust up to quite strong amplitude fuctuations \cite{Carleo2013}.

\end{document}